\documentstyle[12pt]{article}
\begin{document}
\begin{titlepage}
\begin{flushright}
IC/2001/131\\
hep-th/0110013
\end{flushright}
\vspace{10 mm}

\begin{center}
{\Large Null Geodesics in Brane World Universe}

\vspace{5mm}

\end{center}
\vspace{5 mm}

\begin{center}
{\large Donam Youm\footnote{E-mail: youmd@ictp.trieste.it}}

\vspace{3mm}

ICTP, Strada Costiera 11, 34014 Trieste, Italy

\end{center}

\vspace{1cm}

\begin{center}
{\large Abstract}
\end{center}

\noindent

We study null bulk geodesic motion in the brane world cosmology in the 
RS2 scenario and in the static universe in the bulk of the charged 
topological AdS black hole.  We obtain equations describing the null 
bulk geodesic motion as observed in one lower dimensions.  We find 
that the null geodesic motion in the bulk of the brane world cosmology 
in the RS2 scenario is observed to be under the additional influence of 
extra non-gravitational force by the observer on the three-brane, if the 
brane universe does not possess the ${\bf Z}_2$ symmetry.  As for the 
null geodesic motion in the static universe in the bulk of the charged 
AdS black hole, the extra force is realized even when the brane universe 
has the ${\bf Z}_2$ symmetry.

\vspace{1cm}
\begin{flushleft}
October, 2001
\end{flushleft}
\end{titlepage}
\newpage

\section{Introduction}

Recently, there has been active interest in the possibility that our 
four-dimensional universe might be embedded in higher-dimensional bulk 
spacetime having warped product structure \cite{rs1,rs2}, as such possibility 
may provide possible attractive solutions to the hierarchy and the 
cosmological constant problems.  Unlike the case of the conventional 
compactification with compact extra space, the bulk metrics of the brane 
world scenarios have nontrivial dependence on the extra spatial coordinates.  
This fact implies that the geodesic motion in such bulk spacetime is observed 
in the embedded lower-dimensional spacetime as being under the influence of 
the extra non-gravitational force, called the fifth force 
\cite{gm,cp,kwe,mlw,wml,mwl,wmls,lim,sew,wess}, if the velocity of the test 
particle has non-zero components along the extra spatial directions.  In our 
previous work \cite{youm}, we analyzed bulk geodesic motion in the general 
symmetrically warped spacetime of codimension one and identified the extra 
force.  We found that particle motion due to such force violates conventional 
laws of particle mechanics in lower dimensions, thereby hinting at the 
higher-dimensional origin of the embedded spacetime.  

In this paper, we study null geodesic motion in the bulk of various brane 
universe models.  (Some aspects of geodesics in the brane world scenarios 
have been studied in Refs. \cite{kh,chr,kal,cf,cceh,mvv,ish} and geodesic 
motion in the five-dimensional Kaluza-Klein theory was also previously 
studied in Refs. \cite{sch,kov,get}.)  
In brane world scenarios, gravitons are assumed to propagate 
freely in the bulk, whereas all the matter fields are assumed to be confined 
on the brane.  Therefore, it is expected that gravitons are observed to be 
under the influence of the extra force from the perspective of an observer 
living on the brane.  Furthermore, gravitons with nontrivial motion in the 
extra spatial dimension are observed to be massive from the perspective of the 
lower-dimensional observer.  This can be understood from the fact that the 
momentum component along the extra spatial dimension can be interpreted as 
being related to mass in lower dimensions.  Furthermore, since the extra 
force observed in lower dimensions has nonzero component along the direction 
parallel to the four-velocity, the mass is observed to vary with time.  
These facts may be used to test whether our universe is described by the 
brane world scenario.

The paper is organized as follows.  In section 2, we consider the null 
geodesic motion in the bulk of the brane world cosmology in the 
Randall-Sundrum (RS) scenario  with one positive tension brane and 
noncompact extra space \cite{rs2}, i.e., the RS2 scenario.  We find that 
the extra force is zero (i.e., the null bulk geodesic motion is observed 
as the timelike geodesic motion on the brane) in the ${\bf Z}_2$ symmetric 
brane universe.  The extra force turns out to be proportional to the 
constant parametrizing the extend to which ${\bf Z}_2$ symmetry is broken.  
In section 3, we consider the ${\bf Z}_2$ symmetric static brane 
universe in the bulk of the charged topological AdS black hole.  Even with 
the ${\bf Z}_2$ symmetry, the extra force is observed in lower dimensions.  

\section{Null Bulk Geodesics in the RS2 Brane World Cosmology}

In this section, we study the null bulk geodesic motion in the brane world 
cosmology in the RS2 scenario \cite{rs2}.  The action for the model is given by
\begin{equation}
S=\int d^5x\sqrt{-G}\left[{1\over{2\kappa^2}}{\cal R}-\Lambda\right]+
\int d^4x\sqrt{-g}\left[{\cal L}_{\rm mat}-\sigma\right],
\label{rs2cosact}
\end{equation}
where ${\cal L}_{\rm mat}$ is the Lagrangian for the matter fields on the 
brane, $\kappa$ is the five-dimensional gravitational constant, $\Lambda$ 
is the bulk cosmological constant, and $\sigma$ is the tension of the brane 
assumed to be located at the origin $y=0$ of the extra special coordinate 
$y$.  The metric $g_{\mu\nu}$ on the brane is given in terms of the bulk 
metric $G_{MN}$ by $g_{\mu\nu}(x^{\rho})=G_{\mu\nu}(x^{\rho},0)$.  
The general bulk metric ansatz for the brane world cosmology with stabilized 
extra spatial dimension is
\begin{equation}
G_{MN}dx^Mdx^N=-n^2(t,y)dt^2+a^2(t,y)\gamma_{ij}dx^idx^j+dy^2,
\label{cosmet}
\end{equation}
where $\gamma_{ij}$ is the metric for the maximally symmetric 
three-dimensional space given by
\begin{equation}
\gamma_{ij}dx^idx^j=\left(1+\textstyle{k\over 4}\delta_{mn}x^mx^n\right)^{-2}
\delta_{ij}dx^idx^j.
\label{gammet}
\end{equation}
The curvature parameter $k$ is $-1$, 0, 1, respectively for the 
three-dimensional space with the negative, zero and positive spatial 
curvature.  For the phenomenological relevance, we consider the flat universe 
case ($k=0$), only, in this paper.  

We study the geodesic motion in the bulk spacetime with the metric given 
by Eq. (\ref{cosmet}).  The geodesic motion of a test particle in the bulk 
spacetime is described by the geodesic equations
\begin{equation}
{{d^2x^R}\over{d\lambda^2}}+\hat{\Gamma}^R_{MN}{{dx^M}\over{d\lambda}}
{{dx^N}\over{d\lambda}}=0,
\label{gdseqs}
\end{equation}
which take the following forms after the explicit expression (\ref{cosmet}) 
for the bulk metric with $k=0$ is substituted:
\begin{equation}
{{d^2t}\over{d\lambda^2}}+{\dot{n}\over n}\left({{dt}\over{d\lambda}}\right)^2
+2{n^{\prime}\over n}{{dt}\over{d\lambda}}{{dy}\over{d\lambda}}+{{a\dot{a}}
\over{n^2}}\sum_j\left({{dx^j}\over{d\lambda}}\right)^2=0,
\label{geoeq1}
\end{equation}
\begin{equation}
{{d^2x^i}\over{d\lambda^2}}+2{\dot{a}\over a}{{dt}\over{d\lambda}}{{dx^i}\over
{d\lambda}}+2{a^{\prime}\over a}{{dy}\over{d\lambda}}{{dx^i}\over{d\lambda}}=0,
\label{geoeq2}
\end{equation}
\begin{equation}
{{d^2y}\over{d\lambda^2}}+nn^{\prime}\left({{dt}\over{d\lambda}}\right)^2-
aa^{\prime}\sum_j\left({{dx^j}\over{d\lambda}}\right)^2=0.
\label{geoeq3}
\end{equation}
Here, $\hat{\Gamma}^R_{MN}$ is the Christoffel symbol (of the second kind) for 
the bulk metric $G_{MN}$ and $\lambda$ is an affine parameter for the geodesic 
path $x^M(\lambda)$.  The affine parameter $\lambda$ is defined by the 
following metric compatibility condition along the geodesic path:
\begin{equation}
-\epsilon_5=G_{MN}{{dx^M}\over{d\lambda}}{{dx^N}\over{d\lambda}}=
-n^2\left({{dt}\over{d\lambda}}\right)^2+a^2\delta_{ij}{{dx^i}\over{d
\lambda}}{{dx^j}\over{d\lambda}}+\left({{dy}\over{d\lambda}}\right)^2,
\label{metcpcd}
\end{equation}
where $\epsilon_5=0,1$ respectively for a massless test particle (i.e., a 
null geodesic motion) and a massive test particle (i.e., a timelike geodesic 
motion).  Since it is generally assumed in the brane world scenarios that 
only gravitons can propagate freely in the bulk, in this paper we consider 
the null bulk geodesic motion (i.e., the $\epsilon_5=0$ case), only.  

We rewrite the bulk geodesic equations (\ref{geoeq1}-\ref{geoeq3}) in terms 
of quantities of four-dimensional spacetime on the hypersurface $y=const$, 
for the purpose of studying particle dynamics as observed in one lower 
dimensions.  Since the metric on the hypersurface is given by
\begin{equation}
g_{\mu\nu}dx^{\mu}dx^{\nu}=-n^2dt^2+a^2\delta_{ij}dx^idx^j,
\label{branmet}
\end{equation}
the affine parameter $\tilde{\lambda}$ for the motion observed on the 
hypersurface is defined by
\begin{equation}
-\epsilon_4=g_{\mu\nu}{{dx^{\mu}}\over{d\tilde{\lambda}}}{{dx^{\nu}}\over
{d\tilde{\lambda}}}=-n^2\left({{dt}\over{d\tilde{\lambda}}}\right)^2+a^2
\delta_{ij}{{dx^i}\over{d\tilde{\lambda}}}{{dx^j}\over{d\tilde{\lambda}}},
\label{affnbrnmtn}
\end{equation}
where $\epsilon_4=-1,0,1$ respectively for a spacelike, a lightlike and a 
timelike motions observed on the three-brane.  The $y=0$ case, for which 
$n=1$ and $a=a_0(t)\equiv a(t,0)$, corresponds to the metric compatibility 
condition for the motion observed on the three-brane.  
We assume that the affine parameter $\tilde{\lambda}$ for the motion 
observed on the hypersurface is a smooth function of the affine parameter 
$\lambda$ for the bulk geodesic motion: $\tilde{\lambda}=f(\lambda)$.  Then, 
the consistency equation (\ref{metcpcd}) for the bulk geodesic motion can 
be rewritten in terms of the new parameter $\tilde{\lambda}$ as
\begin{equation}
-n^2\left({{dt}\over{d\tilde{\lambda}}}\right)^2+a^2\delta_{ij}{{dx^i}\over{d
\tilde{\lambda}}}{{dx^j}\over{d\tilde{\lambda}}}=-\left[\epsilon_5\left(
{{d\lambda}\over{d\tilde{\lambda}}}\right)^2+\left({{dy}\over{d
\tilde{\lambda}}}\right)^2\right].
\label{blkthrcv}
\end{equation}
For gravitons ($\epsilon_5=0$), which freely move along the extra spatial 
direction (i.e., ${{dy}\over{d\lambda}}\neq 0$), the bulk geodesic motion 
can be observed only as timelike ($\epsilon_4=1$) on the hypersurface.  
Namely, gravitons are observed to be massive by observers living on the 
hypersurface.  The new parameter $\tilde{\lambda}$ is an affine parameter 
for such motion, if the following is satisfied:
\begin{equation}
\left({{dy}\over{d\tilde{\lambda}}}\right)^2=1,
\label{condfrtl}
\end{equation}
for which the RHS of Eq. (\ref{blkthrcv}) becomes $-1$.  
To express the bulk geodesic equations (\ref{geoeq1}-\ref{geoeq3}) in 
terms of the new parameter $\tilde{\lambda}$, we obtain the relation between 
the two parameters $\lambda$ and $\tilde{\lambda}$ by making use of Eqs. 
(\ref{geoeq3},\ref{condfrtl}).  The relation is given by
\begin{equation}
\left({{d\tilde{\lambda}}\over{d\lambda}}\right)^{-1}{d\over{d\tilde{\lambda}}}
\left({{d\tilde{\lambda}}\over{d\lambda}}\right)=-nn^{\prime}\left({{dt}\over
{d\tilde{\lambda}}}\right)^2+aa^{\prime}\sum_j\left({{dx^j}\over{d\tilde{
\lambda}}}\right)^2.
\label{pararel}
\end{equation}
Making use of Eqs. (\ref{condfrtl},\ref{pararel}), we reexpress the 
$t$- and $x^i$-component bulk geodesic equations (\ref{geoeq1},\ref{geoeq2}) 
in terms of the new parameter $\tilde{\lambda}$ as follows
\begin{equation}
{{d^2t}\over{d\tilde{\lambda}^2}}+{\dot{n}\over n}\left({{dt}\over{d\tilde{
\lambda}}}\right)^2+{{a\dot{a}}\over n^2}\sum_j\left({{dx^j}\over{d\tilde{
\lambda}}}\right)^2=\left[nn^{\prime}\left({{dt}\over{d\tilde{\lambda}}}
\right)^2-aa^{\prime}\sum_j\left({{dx^j}\over{d\tilde{\lambda}}}\right)^2
-2{n^{\prime}\over n}\right]{{dt}\over{d\tilde{\lambda}}},
\label{lwgeod1}
\end{equation}
\begin{equation}
{{d^2x^i}\over{d\tilde{\lambda}^2}}=\left[nn^{\prime}\left({{dt}\over{d
\tilde{\lambda}}}\right)^2-aa^{\prime}\sum_j\left({{dx^j}\over{d\tilde{
\lambda}}}\right)^2-2{a^{\prime}\over a}\right]{{dx^i}\over{d\tilde{\lambda}}}.
\label{lwgeod2}
\end{equation}
Note, the LHS's of these equations have the forms ${{d^2x^{\rho}}\over{d
\tilde{\lambda}^2}}+\Gamma^{\rho}_{\mu\nu}{{dx^{\mu}}\over{d\tilde{\lambda}}}
{{dx^{\nu}}\over{d\tilde{\lambda}}}$, where $\Gamma^{\rho}_{\mu\nu}$ is the 
Christoffel symbol for the metric $g_{\mu\nu}dx^{\mu}dx^{\nu}=-n^2(t,y_0)dt^2+
a^2(t,y_0)\delta_{ij}dx^idx^j$ on the hypersurface $y=y_0=const$.  
The RHS's are identified as the time and the spatial components of 
the four-acceleration vector $A^{\mu}$ of the particle due to 
non-gravitational force.  So, we see that the null bulk geodesic motion 
is observed on the hypersurface to be the timelike motion under the 
additional influence of extra non-gravitational force.  

The observer on the hypersurface will find something unusual about the 
four-acceleration vector.  Namely, the four-acceleration vector has 
nonzero component parallel to the four-velocity:
\begin{equation}
g_{\mu\nu}A^{\mu}{{dx^{\nu}}\over{d\tilde{\lambda}}}=\left[2-n^2\left(
{{dt}\over{d\tilde{\lambda}}}\right)^2+a^2\sum_i\left({{dx^i}\over{d
\tilde{\lambda}}}\right)^2\right]\left[nn^{\prime}\left({{dt}\over{d
\tilde{\lambda}}}\right)^2-aa^{\prime}\sum_i\left({{dx^i}\over{d
\tilde\lambda}}\right)^2\right],
\label{nzrparacmp}
\end{equation}
contrary to the conventional laws of particle mechanics.  
An observer who is unaware of the existence of extra dimensions will 
interpret such abnormal four-acceleration component as being due to 
the variation of the particle mass \cite{mwl,lim} in the following way.  
Taking into account the possibility of time-variable proper mass $m$ 
of a particle, we write the force law for the particle motion in 
curved space time as
\begin{equation}
F^{\mu}={{Dp^{\mu}}\over{d\tilde{\lambda}}}={{dm}\over{d\tilde{\lambda}}}
{{dx^{\mu}}\over{d\tilde{\lambda}}}+mA^{\mu},
\label{frclaw}
\end{equation}
where $p^{\mu}\equiv m{{dx^{\mu}}\over{d\tilde{\lambda}}}$ is the 
four-momentum of the particle.  The four-dimensional observer will assume 
that $g_{\mu\nu}F^{\mu}{{dx^{\nu}}\over{d\tilde{\lambda}}}=0$, since all the 
known forces in nature act perpendicularly to the four-velocity ${{dx^{\mu}}
\over{d\tilde{\lambda}}}$ of a particle.  So, the four-dimensional observer 
will conclude that the mass of the particle varies with time as
\begin{equation}
{1\over m}{{dm}\over{d\tilde{\lambda}}}=g_{\mu\nu}A^{\mu}{{dx^{\nu}}\over
{d\tilde{\lambda}}},
\label{tvarmss}
\end{equation}
which says that the proper mass of a particle varies with time when its 
four-acceleration vector has nonzero component parallel to its four-velocity 
vector.  As was pointed out in Ref. \cite{sew}, the component of the 
four-acceleration vector parallel to the four-velocity vector can be 
removed through non-affine transformation of the parameter $\tilde{\lambda}$, 
and therefore the time variation of the particle mass may be regarded as an 
artifact of choosing wrong parameter $\tilde{\lambda}$ for the particle 
motion.  In fact, there is an ambiguity in choosing the right parameter 
for the particle motion.  However, since the metric on the hypersurface 
is given by $g_{\mu\nu}$, an observer, who is unaware of the extra dimension, 
will choose the parameter $\tilde{\lambda}$ satisfying $g_{\mu\nu}{{dx^{\mu}}
\over{d\tilde{\lambda}}}{{dx^{\nu}}\over{d\tilde{\lambda}}}=-1$ as the 
natural parameter describing a timelike motion observed on the hypersurface.  

We now study the bulk geodesic motion as observed on the three-brane 
(located at $y=0$).  Since the first $y$ derivatives of metric components 
are discontinuous at $y=0$ due to the $\delta$-function singularity 
there, Eqs. (\ref{lwgeod1},\ref{lwgeod2}) are not well-defined at 
$y=0$.  Nevertheless, we can obtain the effective equations on the 
three-brane by taking the mean values of Eqs. (\ref{lwgeod1},
\ref{lwgeod2}) across $y=0$ and applying the following boundary 
conditions on the first derivatives of the metric components:
\begin{equation}
[a^{\prime}]_0=-{\kappa^2\over 3}a_0\left(\sigma+\varrho\right),
\label{ydera}
\end{equation}
\begin{equation}
[n^{\prime}]_0=-{\kappa^2\over 3}\left(\sigma-3\wp-2\varrho\right),
\label{ydern}
\end{equation}
where $\varrho$ and $\wp$ are the mass density and the pressure of the 
brane matter fields, and $[F]_0\equiv F(0^+)-F(0^-)$ denotes the jump of 
$F(y)$ across $y=0$.  We define the mean value of a function $F$ across 
$y=0$ as $\sharp F\sharp\equiv[F(0^+)+F(0^-)]/2$.  
When the brane universe is invariant under the ${\bf Z}_2$ transformation, 
$y\to -y$, then the first derivatives of the metric components satisfy 
$a^{\prime}(0^+)=-a^{\prime}(0^-)$ and $n^{\prime}(0^+)=-n^{\prime}(0^-)$.  
So, the mean values of Eqs. (\ref{lwgeod1},\ref{lwgeod2}) across $y=0$ 
respectively take the forms:
\begin{equation}
{{d^2t}\over{d\tilde{\lambda}^2}}+a_0\dot{a}_0\sum_j\left({{dx^j}\over
{d\tilde{\lambda}}}\right)^2=0,
\label{thrbrneq1}
\end{equation}
\begin{equation}
{{d^2x^i}\over{d\tilde{\lambda}^2}}=0,
\label{thrbrneq2}
\end{equation}
where we made use of the fact that the time coordinate $t$ is defined such 
that $n=1$ at $y=0$.  These are just geodesic equations for a test particle 
moving in the gravitational field $g_{\mu\nu}dx^{\mu}dx^{\nu}=-dt^2+a^2_0(t)
\delta_{ij}dx^idx^j$.  Namely, the null bulk geodesic motion is observed on 
the three-brane as the timelike geodesic motion.  
Next, we consider the brane universe without the ${\bf Z}_2$ symmetry.  
Without the ${\bf Z}_2$ symmetry, the first derivative of $a$ around $y=0$ 
satisfies
\begin{equation}
a^{\prime}(0^+)=-a^{\prime}(0^-)+d(t),
\label{frstansym}
\end{equation}
where $d(t)$ is an arbitrary function of $t$.  By applying the boundary 
condition (\ref{ydera}) and taking the jump of the $(y,y)$-component 
Einstein equation, $d(t)$ can be determined to take the following form 
\cite{ddpv}:
\begin{equation}
d(t)={{2F}\over{(\sigma+\varrho)a^3_0}},
\label{asymd}
\end{equation}
where an integration constant $F$ parametrizes the extent to which the 
${\bf Z}_2$ symmetry is broken.  The mean values of Eqs. (\ref{lwgeod1},
\ref{lwgeod2}) then respectively take the following forms:
\begin{equation}
{{d^2t}\over{d\tilde{\lambda}^2}}+a_0\dot{a}_0\sum_j\left({{dx^j}\over
{d\tilde{\lambda}}}\right)^2=-{F\over{(\sigma+\varrho)a^2_0}}{{dt}\over
{d\tilde{\lambda}}}\sum_j\left({{dx^j}\over{d\tilde{\lambda}}}\right)^2,
\label{brnpeq1}
\end{equation}
\begin{equation}
{{d^2x^i}\over{d\tilde{\lambda}^2}}=-{F\over{\sigma+\varrho}}\left[
{1\over a^2_0}\sum_j\left({{dx^j}\over{d\tilde{\lambda}}}\right)^2+
{2\over a^4_0}\right]{{dx^i}\over{d\tilde{\lambda}}}.
\label{brnpeq2}
\end{equation}
So, the null bulk geodesic motion is observed on the three-brane of 
asymmetric brane universe to be under the additional influence of 
non-gravitational force.  The four-acceleration vector $A^{\mu}$ has 
non-zero component parallel to the four-velocity and therefore the 
proper mass of the particle is observed to vary with time as
\begin{equation}
{1\over m}{{dm}\over{d\tilde{\lambda}}}=g_{\mu\nu}A^{\mu}{{dx^{\nu}}
\over{d\tilde{\lambda}}}={F\over{(\sigma+\varrho)a^2_0}}\left[\left(
{{dt}\over{d\tilde{\lambda}}}\right)^2-a^2_0\sum_i\left({{dx^i}\over
{d\tilde{\lambda}}}\right)^2-2\right]\sum_i\left({{dx^i}\over{d\tilde{
\lambda}}}\right)^2.
\label{chmssasym}
\end{equation}
The four-acceleration of the particle and the time variation of the 
particle mass are proportional to the asymmetry parameter $F$, and therefore 
vanish in the limit of $F=0$, for which the brane world becomes 
${\bf Z}_2$ symmetric.  We expect that these results continue to hold for 
the brane world cosmology in the RS1 model \cite{rs1}.  Making use of these 
facts, we can experimentally determine whether our universe, assumed to be 
modeled by the RS scenario, is ${\bf Z}_2$ symmetric or not.

\section{Null Bulk Geodesics in Static Universe in the bulk of Charged AdS 
Black Hole}

In this section, we consider the brane universe in the bulk of the charged 
topological AdS black hole.  The bulk metric has the following form:
\begin{eqnarray}
G_{MN}dx^Mdx^N&=&-h(y)dt^2+y^2\gamma_{ij}dx^idx^j+{1\over{h(y)}}dy^2,
\cr
h(y)&=&k-{\mu\over y^2}+{q^2\over y^4}+{y^2\over l^2},
\label{asymmet}
\end{eqnarray}
where $q$ and $\mu$ are respectively proportional to the charge and the ADM 
mass of the black hole, and $l$ is the curvature radius of the background 
AdS spacetime.  For the $k=0$ case, the bulk geodesic equations (\ref{gdseqs}) 
take the following forms:
\begin{equation}
{{d^2t}\over{d\lambda^2}}+{h^{\prime}\over h}{{dt}\over{d\lambda}}{{dy}
\over{d\lambda}}=0,
\label{geodeq1}
\end{equation}
\begin{equation}
{{d^2x^i}\over{d\lambda^2}}+{2\over y}{{dx^i}\over{d\lambda}}{{dy}\over
{d\lambda}}=0,
\label{geodeq2}
\end{equation}
\begin{equation}
{{d^2y}\over{d\lambda^2}}+{{hh^{\prime}}\over 2}\left({{dt}\over{d\lambda}}
\right)^2-{1\over 2}{h^{\prime}\over h}\left({{dy}\over{d\lambda}}\right)^2-
hy\sum_j\left({{dx^j}\over{d\lambda}}\right)^2=0,
\label{geodeq3}
\end{equation}
and the metric compatibility condition along the geodesic path becomes
\begin{equation}
-\epsilon_5=-h\left({{dt}\over{d\lambda}}\right)^2+y^2\delta_{ij}{{dx^i}\over
{d\lambda}}{{dx^j}\over{d\lambda}}+{1\over h}\left({{dy}\over{d\lambda}}
\right)^2.
\label{metcmpcd2}
\end{equation}
In terms of a new parameter $\tilde{\lambda}=f(\lambda)$, the compatibility 
condition (\ref{metcmpcd2}) is rewritten as
\begin{equation}
-h\left({{dt}\over{d\tilde{\lambda}}}\right)^2+y^2\delta_{ij}{{dx^i}\over
{d\tilde{\lambda}}}{{dx^j}\over{d\tilde{\lambda}}}=-\left[\epsilon_5\left(
{{d\lambda}\over{d\tilde{\lambda}}}\right)^2+{1\over h}\left({{dy}\over{d
\tilde{\lambda}}}\right)^2\right].
\label{cmpcnd2}
\end{equation}
So, the null bulk geodesic motion ($\epsilon_5=0$) with ${{dy}\over{d
\tilde{\lambda}}}\neq 0$ is observed on the hypersurface $y=const$ as 
timelike.  The parameter $\tilde{\lambda}$ is an affine parameter for 
such motion observed on the hypersurface, if the following is satisfied:
\begin{equation}
\left({{dy}\over{d\tilde{\lambda}}}\right)^2=h,
\label{tlmbcnd}
\end{equation}
for which the RHS of Eq. (\ref{cmpcnd2}) becomes $-1$.  
From Eqs. (\ref{geodeq3},\ref{tlmbcnd}), we obtain the following relation 
between the parameters $\lambda$ and $\tilde{\lambda}$:
\begin{equation}
\left({{d\tilde{\lambda}}\over{d\lambda}}\right)^{-1}{d\over{d\tilde{
\lambda}}}\left({{d\tilde{\lambda}}\over{d\lambda}}\right)=-\sqrt{h}
\left[{h^{\prime}\over 2}\left({{dt}\over{d\tilde{\lambda}}}
\right)^2-y\sum_j\left({{dx^j}\over{d\tilde{\lambda}}}\right)^2\right].
\label{pararel2}
\end{equation}
Making use of this relation, we can rewrite the $t$- and $x^i$-component 
geodesic equations Eqs. (\ref{geodeq1},\ref{geodeq2}) in terms of the 
new parameter $\tilde{\lambda}$ as
\begin{equation}
{{d^2t}\over{d\tilde{\lambda}^2}}=\sqrt{h}\left[{h^{\prime}
\over 2}\left({{dt}\over{d\tilde{\lambda}}}\right)^2-y\sum_j\left({{dx^j}
\over{d\tilde{\lambda}}}\right)^2-{h^{\prime}\over{h}}\right]
{{dt}\over{d\tilde{\lambda}}},
\label{geq1}
\end{equation}
\begin{equation}
{{d^2x^i}\over{d\tilde{\lambda}^2}}=\sqrt{h}\left[{h^{\prime}\over 2}\left(
{{dt}\over{d\tilde{\lambda}}}\right)^2-y\sum_j\left({{dx^j}\over{d\tilde{
\lambda}}}\right)^2-{2\over y}\right]{{dx^i}\over{d\tilde{\lambda}}}.
\label{geq2}
\end{equation}
So, the null bulk geodesic motion is observed on the hypersurface to be a 
timelike motion under the additional influence of non-gravitational 
force.  The four-acceleration vector has non-zero component parallel to 
the four-velocity vector, and therefore the proper mass of the particle 
is observed to vary with time:
\begin{equation}
{1\over m}{{dm}\over{d\tilde{\lambda}}}=g_{\mu\nu}A^{\mu}{{dx^{\nu}}\over
{d\tilde{\lambda}}}=\sqrt{h}\left[2-h\left({{dt}\over{d\tilde{\lambda}}}
\right)^2+y^2\sum_i\left({{dx^i}\over{d\tilde{\lambda}}}\right)^2\right]
\left[{h^{\prime}\over 2}\left({{dt}\over{d\tilde{\lambda}}}\right)^2-
y\sum_i\left({{dx^i}\over{d\tilde{\lambda}}}\right)^2\right].
\label{sectvarm}
\end{equation}

We construct the ${\bf Z}_2$ symmetric brane universe in this bulk 
background.  We introduce a probe three-brane at $y=y_0$ and impose the 
${\bf Z}_2$ symmetry under the transformation $y\to y_0/y^2$ by cutting off 
the spacetime in the region $y\geq y_0$ and then replacing it by a copy of 
the spacetime in the region $y\leq y_0$ transformed under $y\to y_0/y^2$ 
\cite{gro}.  We introduce the brane matter with the mass density $\varrho$ 
and the pressure $\wp$.  In particular, we are interested in static brane 
configuration, where the three-brane remains fixed at $y=y_0$.  Such 
configuration can be achieved by the brane matter satisfying \cite{ceg}
\begin{equation}
6\sqrt{h(y_0)}=\kappa^2\varrho y_0,\ \ \ \ \ \ \ \ \ \ 
18h^{\prime}(y_0)=-\kappa^4\varrho(2\varrho+3\wp)y_0.
\label{statcond}
\end{equation}
Using the explicit expression for $h$ with $k=0$ and assuming the equation 
of state of the form $\wp=\omega\varrho$, we obtain the following relations 
among $y_0$ and the black hole parameters \cite{ceg}:
\begin{equation}
\mu=3\left(l^{-2}+\textstyle{1\over{36}}\kappa^4\omega\varrho^2\right)y^4_0,
\ \ \ \ \ \ \ \ \ 
q^2=2\left(l^{-2}+\textstyle{1\over{72}}\kappa^4(1+3\omega)\varrho^2\right)
y^6_0.
\label{pararels}
\end{equation}
To obtain the equations for the null bulk geodesic motion as observed on the 
three-brane, we just substitute the expressions for $h(y_0)$, $h^{\prime}
(y_0)$ in Eq. (\ref{statcond}) into Eqs. (\ref{geq1},\ref{geq2}).  The 
resulting expressions are given by
\begin{equation}
{{d^2t}\over{d\tilde{\lambda}^2}}=-{\kappa^2\over 6}\varrho\left[{\kappa^4
\over{36}}(2+3\omega)\varrho^2y^2_0\left({{dt}\over{d\tilde{\lambda}}}
\right)^2+y^2_0\sum_j\left({{dx^j}\over{d\tilde{\lambda}}}\right)^2-
2(2+3\omega)\right]{{dt}\over{d\tilde{\lambda}}},
\label{sectexp1}
\end{equation}
\begin{equation}
{{d^2x^i}\over{d\tilde{\lambda}^2}}=-{\kappa^2\over 6}\varrho\left[{\kappa^4
\over{36}}(2+3\omega)\varrho^2y^2_0\left({{dt}\over{d\tilde{\lambda}}}
\right)^2+y^2_0\sum_j\left({{dx^j}\over{d\tilde{\lambda}}}\right)^2-2
\right]{{dx^i}\over{d\tilde{\lambda}}}.
\label{sectexp2}
\end{equation}
So, the time variation of the proper mass as observed on the brane is
\begin{eqnarray}
{1\over m}{{dm}\over{d\tilde{\lambda}}}=g_{\mu\nu}A^{\mu}{{dx^{\nu}}\over
{d\tilde{\lambda}}}&=&-{\kappa^2\over 6}\varrho\left[
2-{\kappa^4\over{36}}\varrho^2y^2_0\left({{dt}\over{d\tilde{\lambda}}}
\right)^2+y^2_0\sum_j\left({{dx^j}\over{d\tilde{\lambda}}}\right)^2\right]
\cr
& &\ \ \ \ \times\left[{\kappa^4\over{36}}(2+3\omega)\varrho^2y^2_0\left(
{{dt}\over{d\tilde{\lambda}}}\right)^2+y^2_0\sum_j\left({{dx^j}\over{d
\tilde{\lambda}}}\right)^2\right].
\label{sectvarm2}
\end{eqnarray}
Unlike the case in the previous section, the observer on the three-brane 
will observe extra non-gravitational force, even when the brane universe 
possesses the ${\bf Z}_2$ symmetry.  The strength of the extra force increases 
as mass density of brane matter increases.

\end{document}